# UCov: a User-Defined Coverage Criterion for Test Case Intent Verification

**RAWAD ABOU ASSI, FADI ZARAKET, and WES MASRI**, American University of Beirut

The goal of regression testing is to ensure that the behavior of existing code, believed correct by previous testing, is not altered by new program changes. We argue that the primary focus of regression testing should be on code associated with: a) earlier bug fixes; and b) particular application scenarios considered to be important by the tester. Existing coverage criteria do not enable such focus, e.g., 100% branch coverage does not guarantee that a given bug fix is exercised or a given application scenario is tested. Therefore, there is a need for a new and complementary coverage criterion in which *the user can define* a *test requirement characterizing a given behavior to be covered* as opposed to choosing from a pool of pre-defined and generic program elements. We propose this new methodology and call it *UCov*, a *user-defined coverage criterion* wherein a test requirement is an *execution pattern* of program elements and predicates. Our proposed criterion is not meant to replace existing criteria, but to complement them as it focuses the testing on important code patterns that could go untested otherwise.

*UCov* supports *test case intent verification*. For example, following a bug fix, the testing team may augment the regression suite with the test case that revealed the bug. However, this test case might become obsolete due to code modifications not related to the bug. But if an execution pattern characterizing the bug was defined by the user, *UCov* would determine that test case intent verification failed. It is also worth mentioning that our methodology paves the way for *test case intent preservation*, e.g., a failed verification could be followed by automated test case generation, the subject of future work.

We implemented our methodology for the Java platform and applied it onto two real life case studies. Our implementation comprises the following: 1) an Eclipse plugin allowing the user to easily specify non-trivial test requirements; 2) the ability of cross referencing test requirements across subsequent versions of a given program; and 3) the ability of checking whether user-defined test requirements were satisfied, i.e., test case intent verification.

Categories and Subject Descriptors: D.2.1 [**Software Engineering**]: Requirements/Specifications - *Tools*; D.2.4 [**Software Engineering**]: Software/Program Verification – *Reliability, Validation*; D.2.5 [**Software Engineering**]: Testing and Debugging - *Testing tools*

Additional Key: software testing; regression testing; validation testing; user-defined coverage criterion; test case intent verification; test case intent preservation.

---

This research was supported in part by NSF (grant# 0819987) and by the Lebanese National Council for Scientific Research.
The original idea behind this work is presented in: "Coverage Specification for Test Case Intent Preservation in Regression Suites" presented at Regression/ICST 2013, Luxembourg, 2013.
Authors' addresses: R. Abou Assi, F. Zaraket, and W. Masri, Electrical and Computer Engineering Department, American University of Beirut, Beirut, Lebanon 1107-2020. Emails: ria21@aub.edu.lb, fz11@aub.edu.lb, and wm13@aub.edu.lb (corresponding author).

## 1. INTRODUCTION

In practice, program correctness is mainly affirmed through *testing*, i.e., by checking that the program produces the expected output. *Regression testing* is an essential part of the maintenance phase of software development; its goal is to ensure that the behavior of existing code, believed correct by previous testing, is not altered by new program changes. Since exhaustive testing is not feasible, testers rely on *coverage criteria* to guide their test selection and to provide a stopping rule for testing.

We argue that the primary focus of regression testing should be on code associated with: a) earlier bug fixes; and b) particular application scenarios considered to be important by the developer or tester. Existing coverage criteria do not enable such focus, e.g., 100% branch coverage does not guarantee that a given bug fix is exercised or a given application scenario is tested. Therefore, there is a need for a new and complementary coverage criterion in which *the user can define* a *test requirement characterizing a given behavior to be covered* as opposed to choosing from a pool of pre-defined and generic program elements. We propose this new methodology and call it *UCov*, a *user-defined coverage criterion* wherein a test requirement [Ammann and Offutt 2008] is an *execution pattern* of program elements and predicates. Our proposed criterion is not meant to replace existing criteria but to complement them as it focuses the testing on important code patterns that could go untested otherwise.

*UCov* supports *test case intent verification*. For example, following a bug fix, the testing team may augment the regression test suite with the test case that revealed the bug. Evidently, this new test case induces an execution pattern associated with the bug; however, it might become obsolete due to code modifications not related to the bug. But our coverage criterion, based on a user defined execution pattern (a test requirement) characterizing the bug and coupled with the test case, would:
   a) Detect whether the test requirement was satisfied or not.
   b) Determine whether test case intent verification passed or failed.
   c) Deem the test suite deficient in case test intent verification failed. Thus, suggesting that a new test case that satisfy the requirement needs to be (manually) generated.

It is also worth mentioning that our approach paves the way for *test case intent preservation*. For example, in the scenario above, a failed verification could be followed by automated test case generation whose aim is to satisfy the user-defined test requirement and thus *preserve* the intent of the test case. This topic will be addressed in future work.

Developers and testers leverage *use case scenarios* when designing test cases. These use case scenarios develop into initial test suites and program implementations. During maintenance, the introduction of new features results in augmenting the test suites with test cases that cover the added features and associated code modifications. The same applies to reported bugs and corresponding fixes. Intuitively, *UCov* documents the relation between the test cases and the corresponding code modifications in a manner that enables test case intent verification and preservation. Currently, the documentation of that relation often exists in the form of modification request records in source control repositories. *UCov* provides an Eclipse plugin to allow the user to express test case intent, i.e., to specify user-defined test requirements using a friendly graphical interface. In future work, we will explore extracting the test requirements automatically from source control repositories.

Current coverage criteria limit the user to choosing from a set of program elements that vary in the level of granularity and complexity. Those include statements, branches, logic expressions [Ammann and Offutt 2003], def-uses [Frankl and Weyuker 1988], dependence chains, predicates, information flow pairs [Masri and Halabi 2011], slice pairs [Masri 2008], and paths [Ball and Larus 1996]. At first, it might appear that what we are proposing is simply to cover more complex test requirements comprised of some patterns or combinations of existing program elements. But in fact, the main goal and contribution of our methodology is to cover *behaviors* as opposed to generic structural program elements, and to couple tests with intents to be verified and preserved. Noting that, to our knowledge, neither of these concepts has been previously proposed, and as Sections 3 and 5 demonstrate, they fill in an important gap lacking in existing coverage criteria.

We implemented our methodology for the Java platform in a tool that provides the following:
   a) *An Eclipse plugin to enable users to easily define test requirements*.
   b) *The ability of cross referencing the test requirements across subsequent versions of a given program*, which is a non-trivial task due to the code differences between versions.
   c) *The ability to determine whether the test requirements are satisfied*, which entails instrumenting the *System Under Test* (*SUT*) at the byte code level.

We applied *UCov* onto two real life case studies; the first case study involves a bug fix, and the second is a scenario of significance to program requirements.

The main advantages of *UCov* to the software maintenance process are described below:
   a) *Bug resurrection happens when faulty code that was fixed, gets introduced again*. Typically this might happen due to the uncoordinated access of a file in a source control system by more than one developer. *UCov* ensures the coverage of the test requirement associated with the bug fix and thus uncovers the resurrecting bug. Without *UCov*, resurrecting bugs might escape typical structural coverage based testing.
   b) *A Bug fix could become faulty due to other code changes* (i.e., a bug was introduced in the bug fix). Here also, *UCov* can detect that the test requirement associated with the bug fix is not satisfied, which calls for revisiting the bug fix and test suite.
   c) In *UCov*, a test case *t* that was coupled with a bug fix, a feature, or some scenario of interest to the tester/developer, is intended to verify an expected (correct) behavior of the application. But if *t* becomes obsolete, that expected behavior would go unverified, which will be detected by *UCov*.
   d) Understandably, even full coverage achieved by existing structural coverage criteria does not establish that all (or any) of the scenarios of a given algorithm are tested. To generalize item c); in *UCov*, each scenario could be coupled with a test case, thus relying on *UCov* to ensure coverage of the scenarios. This enables *validation testing* whose aim is to exercise the functionality of the SU*T*.

We now summarize the contributions of this work:
a. *UCov*, a user-defined coverage criterion for test case intent verification in regression test suites.
b. A methodology that complements existing criteria by focusing the testing on important code patterns that could go untested otherwise. Noting that this capability benefits regression testing as well as validation testing.
c. A methodology that facilitates test case intent preservation.
d. Tool support for the Java platform, downloadable from *webfea.fea.aub.edu.lb/wm13/Research.htm*.

The remainder of this article is organized as follows. Section 2 provides definitions and notation for specifying test requirements. Section 3 motivates the work by walking through three examples. Section 4 describes the main components of *UCov*. Section 5 presents our real life case studies. Section 6 discusses the threats to validity of our methodology. Related work is surveyed in Section 7. Finally, Section 8 presents our future work and conclusions.

## 2. DEFINITIONS AND NOTATION

This section provides definitions for entities relevant to *UCov*, and notation for specifying test requirements.

**Definition -** A *program element* is a basic programming unit such as a statement, a branch, or a definition-use pair.

**Definition -** A *test requirement* is an execution pattern that a *test case must satisfy* or cover.

**Definition –** A *basic test requirement* (*btr*) is a test requirement involving only a set of program elements and a logical expression that describes their execution. For example, basic test requirement $[(s_1 \vee b_1) \wedge (\neg dup_1)]_{btr}$, which involves the set of program elements $\{s_1, b_1, dup_1\}$, is considered to be satisfied if: a) statement $s_1$ or branch $b_1$ did execute, and, b) definition-use pair $dup_1$ did not execute. Note that the logical operators supported by *UCov* are, negation ($\neg$), conjunction ($\wedge$), and disjunction ($\vee$).

**Definition -** A *conditional test requirement* (*ctr*) is a test requirement comprising a test requirement *tr*, and a predicate *p* specifying a state of some program variables. For a conditional test requirement to be satisfied, *tr* should be satisfied, and *p* should evaluate to *true* immediately before. For example, the conditional test requirement $[[s_1 \wedge b_1]_{btr}, x > y]_{ctr}$ requires that statement $s_1$ and branch $b_1$ be executed and, when that happens, *x* be strictly greater than *y*.

**Definition -** A *sequential test requirement* (*str*) is a test requirement composed of a sequence of at least two test requirements that must be satisfied one after the other. For example, the sequential test requirement

[<[$b_1$]$_{btr}$, [$b_2$]$_{btr}$, [$b_3 \vee s_1$]$_{btr}$>]$_{str}$ requires that branches $b_1$ and $b_2$ be sequentially executed, followed by $b_3$ or $s_1$.

***Definition*** - A *repeated test requirement* (*rtr*) is a test requirement comprising a test requirement *tr*, and a range indicating the number of times it should be repeated. For example, the repeated test requirement [[$s_1 \wedge b_1$]$_{btr}$, 5, 1000]$_{rtr}$ requires that statement $s_1$ and branch $b_1$ be executed at least 5 times and at most 1000 times. In case one or both of the bounds do not matter, a "*don't care*" symbol could be specified, e.g., [[$s_1$]$_{btr}$, 100, _]$_{rtr}$ requires that statement $s_1$ be executed at least 100 times.

## 3. MOTIVATION
We now walk through three examples motivating our proposed coverage criterion. The first demonstrates a case involving a bug fix, and the other two involve scenarios of significance.

### 3.1 EXAMPLE 1 – TESTING A BUG FIX
Consider a program $P_1$, an associated test suite $T_1$, and a reported bug that was revealed by $t_{bug}$, a test case not present in $T_1$. The development team fixes the bug to produce $P_2$ and couples $t_{bug}$ with a test requirement that characterizes the bug execution. The testing team augments $T_1$ with $t_{bug}$ to form $T_2$, the regression test suite for $P_2$. Subsequently, $P_2$ is modified to add a feature or to refactor the code, thus, resulting in $P_3$. Assume that the modification renders $t_{bug}$ obsolete as it ceases to satisfy its test requirement. Consequently, $T_2$ becomes inadequate, which calls for replacing $t_{bug}$ with a new test case.

As a concrete example, consider the function *boolean terminateEmployee(int averageSales, int salary)* which determines whether an employee should be terminated or not as follows: a) it computes the next annual raise based on the average sales amount; b) computes the new salary including the raise; and c) recommends termination if the new salary exceeds some threshold (hardcoded to $200,000).

A faulty implementation $P_1$ of *terminateEmployee()* is shown below. The bug is in statement $s_1$ which induces a failure when the computed salary is exactly 200000. An example failing test case would be $t_{bug}$:{(4000000, 170000), false}, where *averageSales* is 4000000, current *salary* is 170000, and the return value is *true* (expected to be *false*). Also, consider test suite $T = \{t_1, t_2, t_3, t_{bug}\}$, where $t_1$:{(1500000, 100000), false}, $t_2$:{(130000, 50000), false}, and $t_3$:{(11000, 35000), false}. Note how $T$ achieves full statement coverage, and contains $t_{bug}$ as the only failing test case.

Due to $t_{bug}$ the developers fix the bug in $P_2$, and couple $t_{bug}$ with a test requirement that characterizes the bug execution, specifically, $t_{bug}$ *is coupled* with $tr_{bug} =$ [<[[$s_1$]$_{btr}$ , *salary == 200000*]$_{ctr}$, [$s_3$]$_{btr}$>]$_{str}$. Meaning, in order for the intent of $tr_{bug}$ to be preserved, salary should have a value of 200000 at $s_1$, and $s_3$ should be executed following it.

Now assume that due to requirements changes, $P_2$ was modified to yield $P_3$. Particularly, two conditional statements were added at the beginning of the function to satisfy the following requirements: 1) if the average sales amount was exceptionally high, do no terminate the employee no matter how high the salary is; and 2) if the average sales amount was exceptionally low, terminate the employee no matter how low the salary is.

These changes have no effect on the execution of $t_1$, $t_2$, or $t_3$, but will render $t_{bug}$ obsolete. That is, the intent of $t_{bug}$ is not preserved in $P_3$ as $tr_{bug}$ is not satisfied anymore. To remedy this problem, which would be alerted by *UCov*, the testing team replaces $t_{bug}$ with $t_{bug}'$: {(2000000, 170000), false} which satisfies $tr_{bug} = [<[[s_1]_{btr}, salary == 200000]_{ctr}, [s_3]_{btr}>]_{str}$. Consequently, the updated test suite becomes $T = \{t_1, t_2, t_3, t_{bug}'\}$.

Furthermore, assume that the bug resurrected in $P_4$, which is not very uncommon in practice. Note how $t_{bug}'$ will reveal the bug in $P_4$. Whereas given a test suite that achieves full coverage will not necessarily do so. For example, test suite $T' = \{t_1, t_2, t_3, t_4, t_5\}$ exhibits 100% statement/branch coverage but does not reveal the bug in $P_4$, where $t_1$:{(1500000, 180000), true}, $t_2$:{(130000, 50000), false}, $t_3$:{(11000, 35000), false}, $t_4$:{(5000000, 150000), false}, and $t_5$:{(900, 20000), false}.

```
// P₁
boolean terminateEmployee(int averageSales, int salary)
{
        int raise = 0;
        if (averageSales >= 1000000) {
                raise = 30000;
        } else if (averageSales >= 100000) {
                raise = 10000;
        } else if (averageSales >= 10000) {
                raise = 1000;
        }
        salary = salary + raise;

s1:     if (salary >= 200000) {  //Bug: should be if (salary > 200000)
s2:             return true;
        }
        else {
s3:             return false;
        }
}
```

```
// P₂
boolean terminateEmployee(int averageSales, int salary)
{
        int raise = 0;
        if (averageSales >= 1000000) {
                raise = 30000;
        } else if (averageSales >= 100000) {
                raise = 10000;
        } else if (averageSales >= 10000) {
                raise = 1000;
        }
        salary = salary + raise;

s1:     if (salary > 200000) {  //Bug is fixed
s2:             return true;
        }
        else {
s3:             return false;
        }
}
```

```
// P₃
boolean terminateEmployee(int averageSales, int salary)
{
        if (averageSales > 3000000) return false; // Added code
        if (averageSales < 1000) return true;    // Added code

        int raise = 0;
        if (averageSales >= 1000000) {
                raise = 30000;
        } else if (averageSales >= 100000) {
                raise = 10000;
        } else if (averageSales >= 10000) {
                raise = 1000;
        }
        salary = salary + raise;

s1:     if (salary > 200000) {
s2:             return true;
        }
        else {
s3:             return false;
        }
}
```

```
// P₄
boolean terminateEmployee(int averageSales, int salary)
{
        if (averageSales > 3000000) return false;
        if (averageSales < 1000) return true;

        int raise = 0;
        if (averageSales >= 1000000) {
                raise = 30000;
        } else if (averageSales >= 100000) {
                raise = 10000;
        } else if (averageSales >= 10000) {
                raise = 1000;
        }
        salary = salary + raise;

s1:     if (salary >= 200000) {  // Resurrected bug
s2:             return true;
        }
        else {
s3:             return false;
        }
}
```

### 3.2 EXAMPLE 2 – TESTING SCENARIOS OF AN ALGORITHM

Typically, algorithms are presented while stressing the prime scenarios they support, which we believe should all be tested for quality assurance. Noting that even full coverage achieved by existing structural coverage criteria does not establish that all (or any) of the scenarios of an algorithm are tested, we advocate our user-defined coverage criterion as an effective solution to this task. Intuitively, each documented scenario (or case) associated with the algorithm describes at least one execution pattern that should be coupled with designated test cases. We illustrate the usage of *UCov* in testing the algorithm for deleting a node in a binary search tree.

The algorithm in Figure 1 presented in Cormen et al. [2001] considers four cases concerning the node $z$ to be deleted:

> *Case1* If $z$ has no children, then it is replaced by NIL.
>
> *Case2* If $z$ has only one child, then it is replaced by that child.
>
> *Case3* If $z$ has two children, then it is replaced by its successor, which is the leftmost node in the sub-tree rooted at the right child of $z$. In this case, the successor of $z$ (say $y$) has no right child. That is, $y$ would be a leaf and thus deleting $z$ would be achieved by replacing the contents of $z$ by those of $y$ and replacing $y$ with NIL.
>
> *Case4* Similarly to *Case3*, $z$ has two children, and is replaced by its successor. However, here $y$ has a right child, and the contents of $z$ are replaced by those of $y$ but instead of replacing $y$ with NIL, it is replaced by its right child.

Figure 2 depicts a test suite $T$ comprising the four test cases $t_1$, $t_2$, $t_3$, and $t_4$. Table 1 details the individual and cumulative statement and branch coverage information for each of the test cases. As shown, $T$ achieves 100% statement coverage and 100% branch coverage.

The execution patterns associated with each of the algorithm's scenarios are also shown at the bottom of Table 1, along with $T$'s coverage information. Test cases $t_1$ and $t_2$ cover the execution patterns (test requirements) of *Case1* and *Case2*, respectively. And both $t_3$ and $t_4$ cover the execution pattern of *Case3*. Therefore, *Case4* is left untested, i.e., none of the tests cover test requirement $[<[s_3]_{btr}, [s_6]_{btr}, [s_8]_{btr}>]_{str}$.

This example demonstrates how applying our coverage criterion would deem test suite $T$ deficient despite the fact that it satisfied full statement and branch coverage. In order to test all four scenarios using *UCov*, the user would: 1) specify their four respective test requirements shown at the bottom of Table 1; and 2) for each test requirement, design at least one test case that satisfies it.

```
BST-DELETE(T, z)
Input: Binary Search Tree (T), pointer to the node to be deleted (z)
Output: Binary Search Tree (T') obtained from T by deleting z

1.      if left[z] = NIL or right[z] = NIL
2.          then y ← z
3.      else y ← TREE-SUCCESSOR(z)

4.      if left[y] ≠ NIL
5.          then x ← left[y]
6.      else x ← right[y]

7.      if x ≠ NIL
8.          then p[x] ← p[y]

9.      if p[y] = NIL
10.         then root[T] ← x
11.     else if y = left[p[y]]
12.         then left[p[y]] ← x
13.     else right[p[y]] ← x

14.     if y ≠ z
15.         then key[z]← key[y]
16.             copy y's satellite data into z
```

**Figure 1-** Pseudo-code for deleting a node in a BST

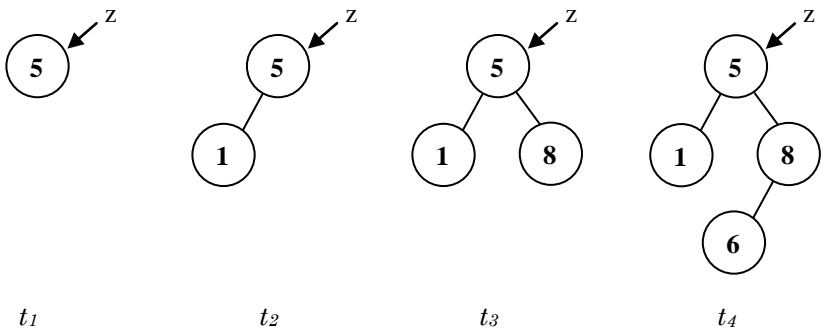

**Figure 2** – Test suite $T = \{t_1, t_2, t_3, t_4\}$

|  |  | $t_1$ | $t_2$ | $t_3$ | $t_4$ | { $t_1$, $t_2$, $t_3$, $t_4$} |
|---|---|---|---|---|---|---|
| **Statements** | S1 | ✓ | ✓ | ✓ | ✓ | ✓ |
|  | S2 | ✓ | ✓ | ✗ | ✗ | ✓ |
|  | S3 | ✗ | ✗ | ✓ | ✓ | ✓ |
|  | S4 | ✓ | ✓ | ✓ | ✓ | ✓ |
|  | S5 | ✗ | ✓ | ✗ | ✗ | ✓ |
|  | S6 | ✓ | ✗ | ✓ | ✓ | ✓ |
|  | S7 | ✓ | ✓ | ✓ | ✓ | ✓ |
|  | S8 | ✗ | ✓ | ✗ | ✗ | ✓ |
|  | S9 | ✓ | ✓ | ✓ | ✓ | ✓ |
|  | S10 | ✓ | ✓ | ✗ | ✗ | ✓ |
|  | S11 | ✗ | ✗ | ✓ | ✓ | ✓ |
|  | S12 | ✗ | ✗ | ✗ | ✓ | ✓ |
|  | S13 | ✗ | ✗ | ✓ | ✗ | ✓ |
|  | S14 | ✓ | ✓ | ✓ | ✓ | ✓ |
|  | S15 | ✗ | ✗ | ✓ | ✓ | ✓ |
|  | S16 | ✗ | ✗ | ✓ | ✓ | ✓ |
| **Branches** | S1→S2 | ✓ | ✓ | ✗ | ✗ | ✓ |
|  | S1→S3 | ✗ | ✗ | ✓ | ✓ | ✓ |
|  | S4→S5 | ✗ | ✓ | ✗ | ✗ | ✓ |
|  | S4→S6 | ✓ | ✗ | ✓ | ✓ | ✓ |
|  | S7→S8 | ✗ | ✓ | ✗ | ✗ | ✓ |
|  | S7→S9 | ✓ | ✗ | ✓ | ✓ | ✓ |
|  | S9→S10 | ✓ | ✓ | ✗ | ✗ | ✓ |
|  | S9→S11 | ✗ | ✗ | ✓ | ✓ | ✓ |
|  | S11→S12 | ✗ | ✗ | ✗ | ✓ | ✓ |
|  | S11→S13 | ✗ | ✗ | ✓ | ✗ | ✓ |
|  | S14→S15 | ✗ | ✗ | ✓ | ✓ | ✓ |
|  | S14→END | ✓ | ✓ | ✗ | ✗ | ✓ |
| **Prime Scenarios** | **Execution Patterns (*TR*)** |  |  |  |  |  |
| *Case1* | $[<[s_2]_{btr}, [s_6]_{btr}, [[s_7]_{btr}, x==NIL]_{ctr}>]_{str}$ | ✓ | ✗ | ✗ | ✗ | ✓ |
| *Case2* | $[<[s_2]_{btr}, [s_8]_{btr}>]_{str}$ | ✗ | ✓ | ✗ | ✗ | ✓ |
| *Case3* | $[<[s_3]_{btr}, [s_6]_{btr}, [[s_7]_{btr}, x==NIL]_{ctr}>]_{str}$ | ✗ | ✗ | ✓ | ✓ | ✓ |
| *Case4* | $[<[s_3]_{btr}, [s_6]_{btr}, [s_8]_{btr}>]_{str}$ | ✗ | ✗ | ✗ | ✗ | ✗ |

**Table 1** – Coverage information for test suite *T*

### 3.3 EXAMPLE 3 – TESTING INACTIVE CLAUSES

This example demonstrates the utility of *UCov* in testing a scenario involving inactive clauses. The scenario discussed here is described in Ammann and Offutt [2008]. Consider the function *bool reset()* in Figure 3 that is designed to control the shutdown system in a nuclear reactor. When the system is in *"override"* mode, the state of a particular valve *("open"* vs. *"closed"*) should not affect the decision to reset the system. A conservative approach would require testing *reset()* in override mode for both positions of the valve. Using *UCov*, this could be achieved by satisfying the following two test requirements:

$[<[[s_1]_{btr}, override==true \land valveClosed==true]_{ctr}, [[s_4]_{btr}, result==true]_{ctr}>]_{str}$
$[<[[s_1]_{btr}, override==true \land valveClosed==false]_{ctr}, [[s_4]_{btr}, result==true]_{ctr}>]_{str}$

```
boolean reset()
{
s1:     boolean result = false;
s2:     if (override || valveClosed)
s3:             result = true;
s4:     return result;
}
```

**Figure 3** – Function to control the shutdown system in a reactor

On the other hand, achieving full statement and branch coverage does not necessarily demonstrate that clause *valveClosed* is inactive, e.g., as when using the following two tests: {*override=true*, $valveClosed=false$} and {*override=false*, $valveClosed=false$}.

## 4. METHODOLOGY AND IMPLEMENTATION

*UCov* entails three main tasks and associated components that we describe next.

### 4.1 SPECIFYING TEST REQUIREMENTS

We first designed and built a programming interface that enables the user to specify test requirements of the types we described in Section 2. Note that our implementation expects the program elements to be specified at the Java bytecode level. Since such interface is only adequate for users who are also developers, we built a graphical Eclipse plugin that hides its complexity, which is downloadable from *webfea.fea.aub.edu.lb/wm13/Research.htm*. The output of the plugin is compilable code that specifies the user-defined test requirements using calls to the programming interface.

The *user's manual* provided with the plugin illustrates in detail how a test requirement is specified graphically. For space limitation we will only show and discuss snapshots from the plugin when presenting the case study in Section 5.1.

Figures 4.a-4.c present the class diagram of the programming interface. A set *TR* of user-defined test requirements comprises a list of basic test requirements (*btr*'s), conditional test requirements (*ctr*'s), sequential test requirements (*str*'s), and repeated test requirements (*rtr*'s). Figure 4.a shows that: 1) the *str*'s, and *rtr*'s are made up of any of the four types of test requirements; and 2) the *ctr*'s are made up of any of the four types of test requirements in addition to a *predicate*. Figure 4.b shows that a *btr* is composed of the conjunction, disjunction, and negation of primitive *btr*'s, which in turn are made up of statements, def-use pairs, and branches. Finally, Figure 4.c shows that a *predicate* is made up of the conjunction, disjunction, and negation of primitive predicates or clauses.

To illustrate the use of *UCov*'s programming interface, the test requirement $[[s_4]_{btr}, result==true]_{ctr}$ associated with function reset shown in Figure 3 would be specified as follows:

```
Statement s1 = new Statement("Reactor", "reset", "()Z", 19);
btr btr1 = new Primitive_btr(s1);
Variable var1 = new Variable("Local", "result", "Reactor", "reset", "()Z");
Predicate pred1 = new Primitive_Predicate("Equal", var1, new Boolean(true));
ctr ctr1 = new ctr(btr1, pred1);
```

The above code assumes that method reset is in class Reactor and $s_4$ is at bytecode offset 19 from the start of reset.

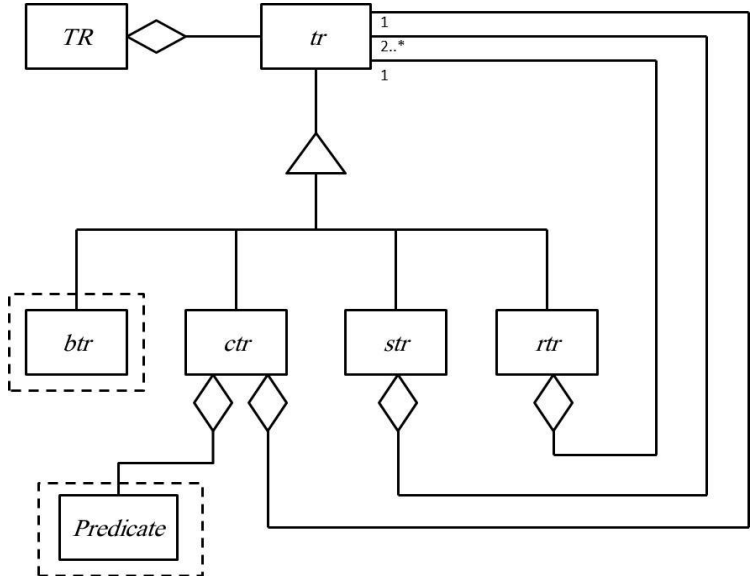

**Figure 4.a)** – Class Diagram of the programming interface

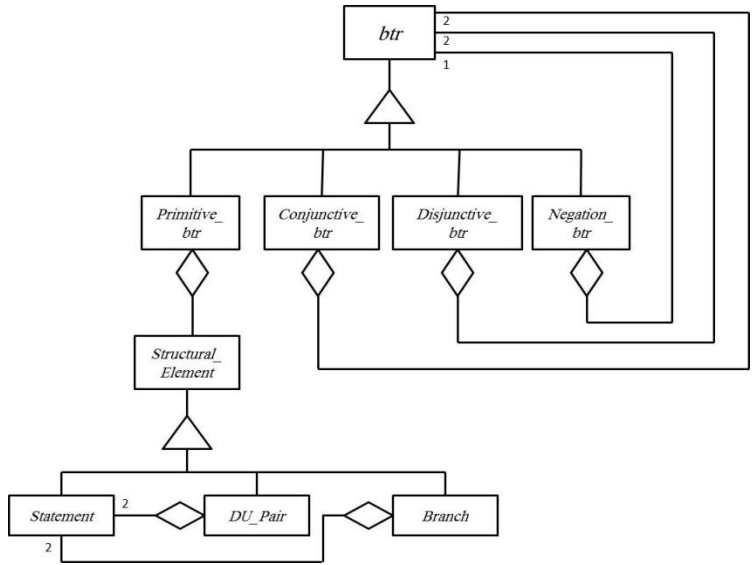

**Figure 4.b)** – Class Diagram associated with *btr* class

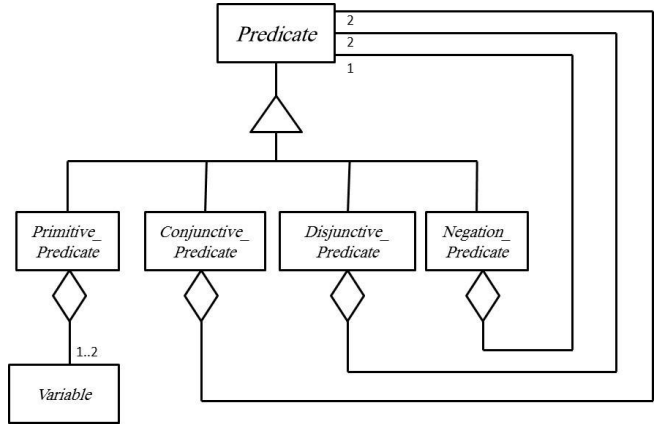

**Figure 4.c)** – Class Diagram associated with *Predicate* class

### 4.2 CROSS REFERENCING TEST REQUIREMENTS ACROSS VERSIONS

The test requirements considered in *UCov* are built from statements, branches, def-uses, and predicates. Since branches and def-uses are constructed from statements, and predicates are constructed from program variables, the task of migrating test requirements across versions boils down to cross referencing statements and variables across versions.

*4.2.1 Statement Mapping*

As our implementation targets the Java platform, we opt to match bytecode statements across versions using a technique inspired from the notion of *Abstract Syntax Tress* [Yang 1991; Fluri et al. 2007; Baxter et al. 1998; Neamtiu et al. 2005]. Given a bytecode statement *s* defined relative to the start of a method *M* in a particular version of the software being considered, our technique identifies the counterpart of *s* relative to the start of *M* in a subsequent version by analyzing what we call the *bytecode dependence tree* (*BDT*) of *M* in both versions.

The *BDT* of a particular function is constructed statically from its list of bytecode instructions $\{s_1, s_2, …, s_n\}$ as follows:
- The tree has *n+1* nodes: a root node labeled "start" and *n* descendant nodes each of which corresponds to one of the bytecode instructions.
- A node *n* is the parent of another node *n'* if one of the following holds:
  - *n* and *n'* respectively correspond to bytecode instructions $s_i$ and $s_j$ such that $s_i$ consumes a value (from the JVM's operand stack) that was produced by $s_j$. This captures the *direct data dependence* relationship described in Masri and Podgurski [2009].
  - *n* is either the "start" node or a node corresponding to a conditional instruction and *n'* represents a non-producer instruction in the direct scope of *n*. This captures the *direct control dependence* relationship described in Masri and Podgurski [2009].
- Siblings are ordered according to their relative positions in the bytecode instruction list.

Figure 5 illustrates the above by showing a snippet of Java code, the corresponding bytecode instructions, and the resulting *BDT*. The nodes of the *BDT* are annotated with the offsets of the corresponding bytecode instructions.

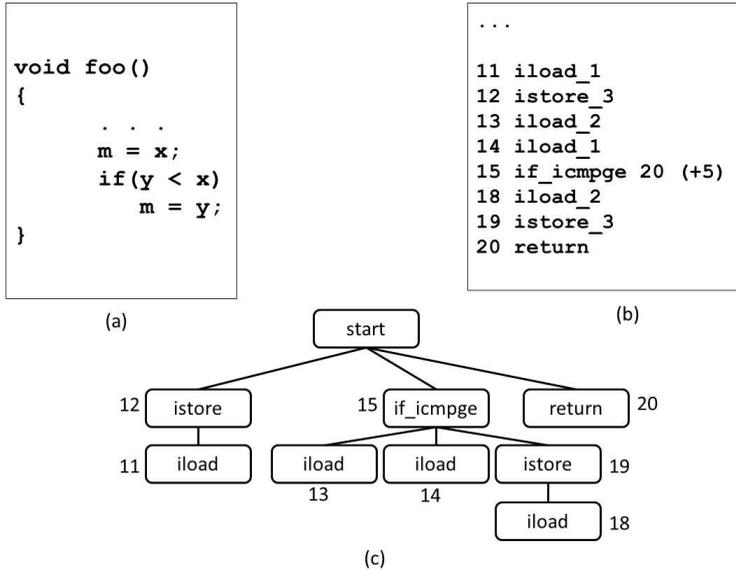

**Figure 5** – (a) Sample method `foo`, (b) Bytecode list of `foo`, (c) *BDT* of `foo`

When trying to match a test requirement *tr* against a subsequent version, every statement *s* in *tr* is mapped. We first check if the code of the method corresponding to *s* (say *M*) has changed between the two versions. If so, we construct the *BDT* of *M* with respect to the original version (say *B*) and that corresponding to the subsequent version (say *B'*). Then, we determine the node in *B'* that is structurally most similar to *s* in *B* using an iterative algorithm as follows:

1. We start with a set of potential candidates. These are the nodes in *B'* whose corresponding bytecode instruction opcode is equal to that of *s*.
2. We repeatedly eliminate the candidates which fail a similarity test of increasing precision. The order we follow is: level-1 descendants, level-1 ancestors, level-2 descendants, level-2 ancestors, level-3 descendants etc. That is, in the first iteration, we eliminate all the candidates whose children (i.e. level-1 descendants) in *B'* are not similar to the children of *s* in *B*. In the second iteration, we eliminate (from the remaining candidates) those whose parents (i.e. level-1 ancestors) in *B'* aren't similar to the parent of *s* in *B*, and so on. And if more than one candidate still remain; we consider the siblings of *s*.
3. The algorithm successfully stops when only one highly similar candidate remains.
4. If at some iteration the set of candidates becomes empty, we restore the results of the previous iteration and require the intervention of the user to resolve the ambiguity. We also require user intervention in case we reached the final iteration with several candidates. However, both scenarios are unlikely to occur.

To demonstrate our mapping mechanism, we consider an "updated" version of the method `foo` of Figure 5. In the new version, shown in Figure 6 with its corresponding bytecode and *BDT*, `foo` is modified by adding a statement that computes the sum of `x` and `y`. In addition, variable `m` is renamed to `min`, to be revisited. We will denote the *BDT* of Figure 5 by *B* and that of Figure 6 by *B'*. Also, we will identify every node using its offset relative to the corresponding *BDT*, (for example *B-11* refers to node 11 in *B*). In what follows, we show how our algorithm maps *B-19* to *B'-29*, i.e., "m = y" in *B* to "min = y" in *B'*. We start with the set of all potential candidates; these are the nodes in *B'* associated with an *istore* instruction, the type of instruction *B-19* is associated with. Therefore, the initial set of candidates consists of *B'-19*, *B'-22*, and *B'-29*. In the first iteration, *B'-19* is eliminated because its child differs from that of *B-19*. Alternatively, *B'-22* and *B'-29* are kept as they pass this first similarity test. In the second iteration, these two candidates undergo the second similarity test which compares their parents with that of *B-19*. As a result, *B'-22* is eliminated and *B'-29* is left as the only candidate. As such, the algorithm terminates by mapping *B-19* to *B'-29*.

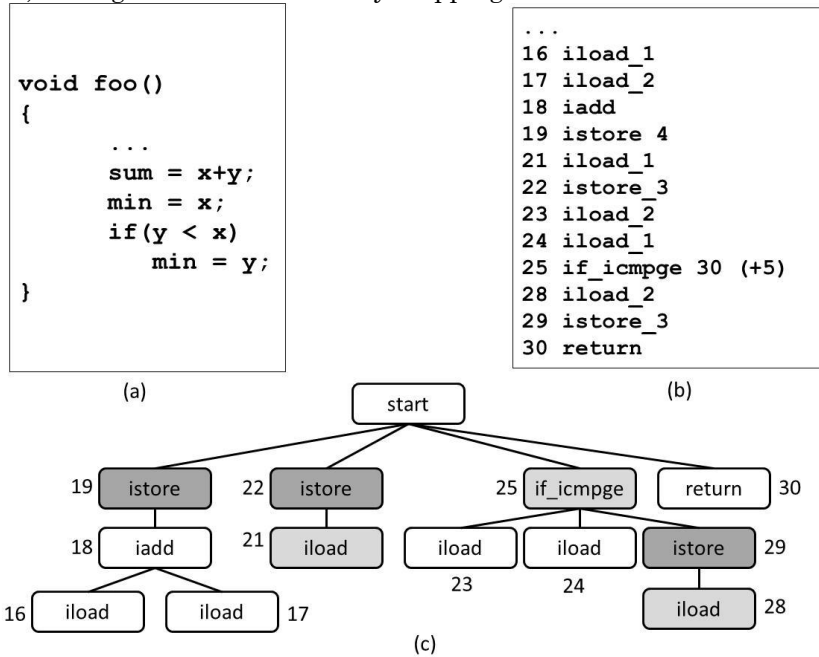

**Figure 6** – Updated version of `foo`

*4.2.1 Variable Mapping*

Our cross referencing technique accounts for variable matching as well. The need for this kind of matching arises when the name of a variable involved in a test requirement is changed in the subsequent version. We leverage the statement mapping mechanism described above as a basis for variable mapping as follows:
1. For each variable *v* to be mapped, we identify the set of bytecode statements referencing it in the original version, say $S=\{s_1, s_2, …, s_k\}$.

2. Then, we perform statement mapping to get the set $S'=\{s_1', s_2', \ldots, s_k'\}$ relevant to the subsequent version.
3. For each $s_i'$, we identify the variable it references and then we consider the counterpart of $v$ to be the variable referenced by all statements in $S'$.
4. As described in step 3, a "perfect" match occurs when all the statements in $S'$ reference the same variable. But if this was not the case, user intervention will be required in order to update the test requirement.

As an example of variable mapping, consider method `foo` and its updated version shown in Figures 5 and 6, respectively. The fact that variable `m` was renamed to `min` (and it is involved in a test requirement) necessitates applying the variable mapping algorithm described above. We first identify the nodes that reference `m` in $B$, which are *B-12* and *B-19*. Applying the statement mapping procedure, we map *B-12* to *B'-22* and *B-19* to *B'-29*. Then, we determine the variable(s) referenced by *B'-22* and *B'-29*. In this case, both nodes reference variable `min`, meaning that the algorithm was successful at perfectly matching `m` and `min`.

### 4.3 CHECKING THE COVERAGE OF TEST REQUIREMENTS

Our approach for checking the coverage of test requirements is to some extent similar to what we adopted in Masri et al. [2013] for the purpose of matching attack signatures. The approach entails two steps: *instrumentation* and *matching*, both of which are done at run-time. For a given program $P$ associated with a set of user-defined test requirements *UTR*, the instrumentation module applies dynamic instrumentation at class load time on $P$ to enable the online matching of the test requirements specified in *UTR*. We implemented our module using the *java.lang.instrument* package, which enables *dynamic* instrumentation, and can be used in conjunction with the bytecode manipulation library *BCEL* [2003], which provides functionality to inject bytecode instructions. The instrumentation package leverages Java *agents*, which are pluggable libraries embedded in the Java Virtual Machine (JVM) that intercept the class-loading process. The agents are run in tandem with the target application and are programmed to carry out the instrumentation. The instrumentation is done by inserting method calls to the matching module at specific locations in $P$. These locations include:
1. Every statement specified in *UTR*. Note that, in case *UTR* was specified in a previous version, statements are mapped according to the approach discussed in Section 4.2.
2. The entry statement of each method specified in *UTR*.
3. Every basic block (*BB*) leader in the method containing a branch specified in *UTR*. A branch entails a source *BB* and a target *BB* both of which belonging to the same method. Instrumenting the source and target of the branch is not sufficient as it is not always the case that the target will execute right after the source. Therefore, all *BB's* in the method must be instrumented in order to track the last executed block. In this manner, for proper branch matching, if the target of the branch is matched, the matcher must check that the last executed *BB* was actually the source.

4. The definition and use statements of each def-use pair (*DUP*) specified in *UTR* as well as all statements that define the variable involved. *DUP'*s form a relationship between a store and a load of a particular variable. For *DUP'*s involving local variables, this relationship is intra-procedural, whereas it might be intra- or inter-procedural for static variables, instance fields, or array elements. *DUP*($s1$, $s2$) signifies that the variable loaded at the use site $s2$ was defined at the given definition site $s1$. Any *killing* definition executing in between these two sites would nullify the definition statement $s1$. Therefore, examining the definition and use locations specified by the test requirement is insufficient, and the instrumentation must treat all other definition sites of the variable involved so as to detect the occurrence of a redefinition of the variable. Instrumenting all possible definition sites entails injecting instructions in all the methods where the variable is defined - except for the case of local variables where only a single method is concerned - naturally adding to the instrumentation overhead; nevertheless, this measure is unavoidable for correctness.

The matching module, on the other hand, keeps track of all the *btr*'s specified in *UTR* as independent test requirements or as part of more complex ones. For every such *btr*, the matching module also maintains a *timestamp* and a *counter* indicating the last time and the number of times it got executed, respectively. In case *UTR* contains *ctr*'s, the matching module would keep track of the "current" values of all involved variables. The matching module is triggered in two cases: 1) state update notification; and 2) structural notification. The first occurs when a variable relevant to *UTR* gets updated. In this case, the value of the corresponding variable is simply updated. The second case occurs when a *btr* referenced by *UTR* gets executed by the program. Here, the matcher updates the timestamp and the counter of the corresponding *btr* and checks all relevant test requirements.

## 5. CASE STUDIES
We now present two real life case studies in which *UCov* is applied.

### 5.1 TESTING A BUG FIX
This case study involves two versions of *NanoXML*, an *XML* parser comprising 7,646 lines of code. The two versions were downloaded along with their test suites from the *SIR* repository (*sir.unl.edu*) and they correspond to versions 1 and 3 in *SIR*. Hereafter, we will refer to these versions as *NanoXML_v1* and *NanoXML_v3*.

A typical *NanoXML* test case involves running a java test program that takes in a certain *XML* file as input and applies some *NanoXML* functionalities on it. Specifically, the test program in our case study is `Parser1_vw_v1.java` and the input file is `testvw_29.xml` shown in Figures 7 and 8, respectively. Basically, The program parses the input file using the `parse()` method defined in `StdXMLParser.java` in the *NanoXML* package and outputs the result.

This test case reveals one of the bugs in *NanoXML_v1* which is fixed in *NanoXML_v3*, namely, a `while` replaced by an `if` in method

*elementAttributesProcessed* in *NonValidator.java*, shown in Figure 9. Figure 10 contrasts the faulty output against the expected output.

```java
public class Parser1_vw_v1
{
    public static void main(String args[]) throws Exception
    {
        if (args.length == 0) {
            System.err.println("Usage: java Parser1_vw_v0 file.xml");
            Runtime.getRuntime().exit(1);
        }
        String filename = args[0];
        IXMLParser parser = XMLParserFactory.createDefaultXMLParser();
        IXMLReader reader = StdXMLReader.fileReader(filename);

        parser.setReader(reader);

        XMLElement xml = (XMLElement) parser.parse();
        (new XMLWriter(System.out)).write(xml);
    }
}
```

**Figure 7** – Test program Parser1_vw_v1.java

```xml
<!DOCTYPE FOO [
    <!ENTITY % extParamEntity SYSTEM "E:\Nanoxml\inputs\nano1\include.ent">
    <!ENTITY value "%extParamEntity;">
    <!ELEMENT FOO (#PCDATA)>
    <!ATTLIST FOO
        x CDATA #REQUIRED
        y CDATA #FIXED "fixedValue"
        z CDATA "defaultValue">
]>

<FOO x='1'>
<VAZ>vaz</VAZ>&value;</FOO>
```

**Figure 8** – Input file testvw_29.xml

```java
public void elementAttributesProcessed(String     name,
                                       String nsPrefix,
                                       String nsSystemId,
                                       Properties extraAttributes)
{
  Properties props = (Properties) this.currentElements.pop(); //s1
  Enumeration _enum = props.keys();                           //s2
  if (_enum.hasMoreElements()) //s3 -- should be while(_enum.hasMoreElements())
  {
    String key = (String) _enum.nextElement();                //s4
    extraAttributes.put(key, props.get(key));                 //s5
  }
}
```

**Figure 9** – Faulty code in NonValidator.java

| `<FOO x="1" z="defaultValue">` | `<FOO x="1" z="defaultValue" y="fixedValue">` |
|---|---|
| `<VAZ>vaz</VAZ>` | `<VAZ>vaz</VAZ>` |
| INCLUDE | INCLUDE |
| `</FOO>`           a) | `</FOO>`                     b) |

**Figure 10** – (a) Faulty output; and (b) Expected output.

Note that the bug fix would be exercised in *NanoXML_v3* only if the **while** loop iterates twice or more. This behavior could be captured in *UCov* by the repeated test requirement $[[s_4]_{btr}, 2, \_]_{rtr}$, or the sequence test requirement $[<[s_4]_{btr}, [s_4]_{btr}>]_{str}$, and possibly others.

We applied *UCov* to capture that behavior and enable testing the bug fix in future releases. Figures 11-14 show snapshots of the *UCov* plugin when used to specify the test requirement $[<[s_4]_{btr}, [s_4]_{btr}>]_{str}$. Specifically, in Figure 11 the user selects a snippet of code corresponding to $s_4$ and *UCov* accordingly lists the bytecode instructions and variables involved in that selection; here we are only concerned with the *astore* instruction, denoted as Statement0. Figure 12 shows how the basic test requirement BTR0 is specified based solely on Statement0. Figure 13 illustrates how the sequence test requirement $[<[s_4]_{btr}, [s_4]_{btr}>]_{str}$, denoted as STR0, is specified in terms of BTR0. And Finally, Figure 14 shows the generated code comprising calls to the programming interface.

Note that in case we opted to specify $[[s_4]_{btr}, 2, \_]_{rtr}$ instead, the steps in Figures 11 and 12 would remain the same and the counterparts of Figures 13 and 14 would be Figures 15 and 16, respectively.

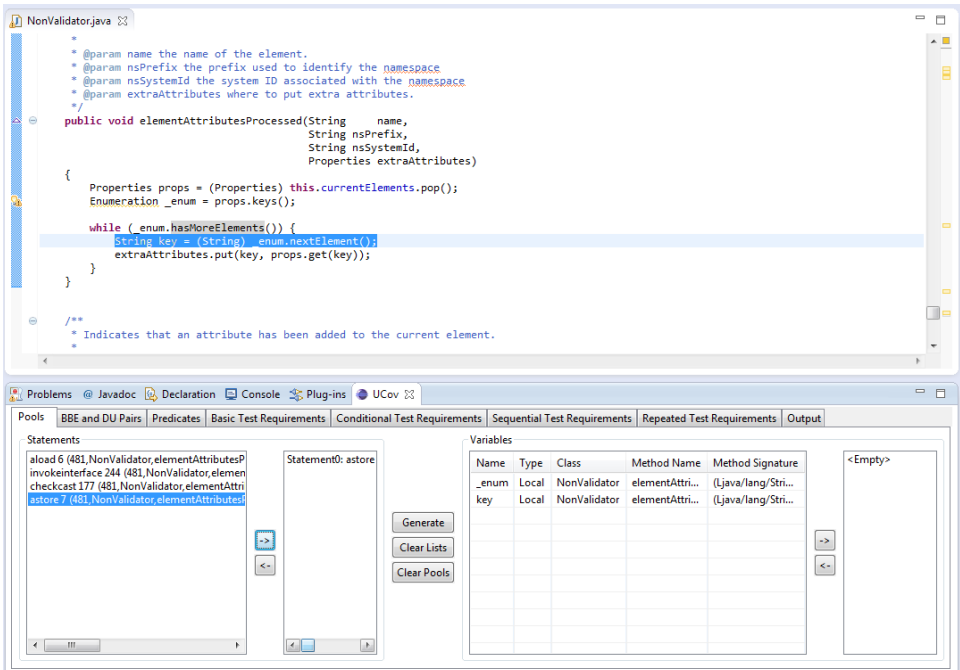

**Figure 11** – Code selection in the *UCov* plugin

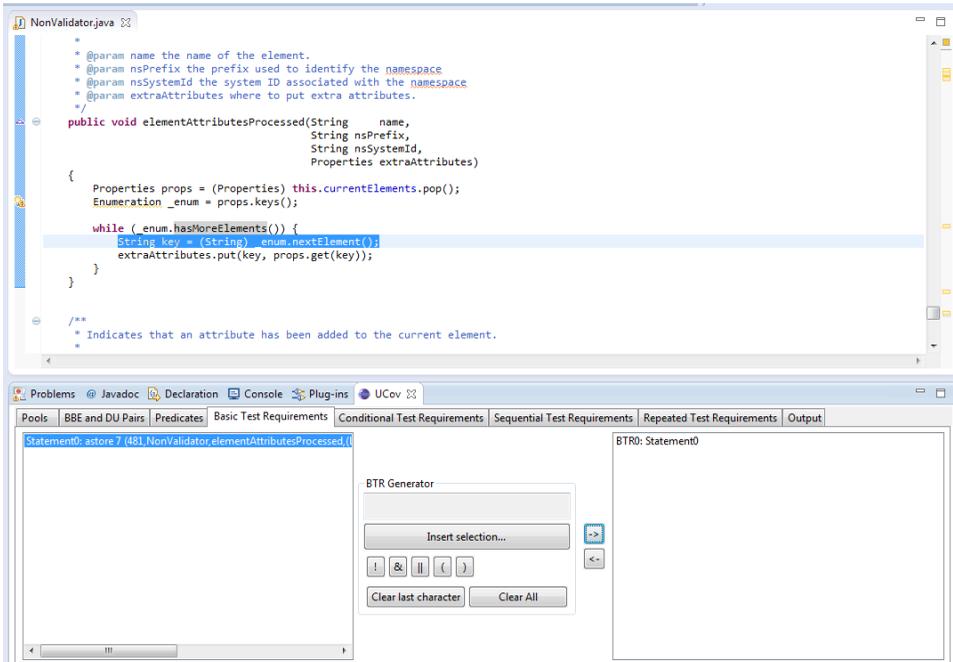

**Figure 12** – *btr* specification in the *UCov* plugin

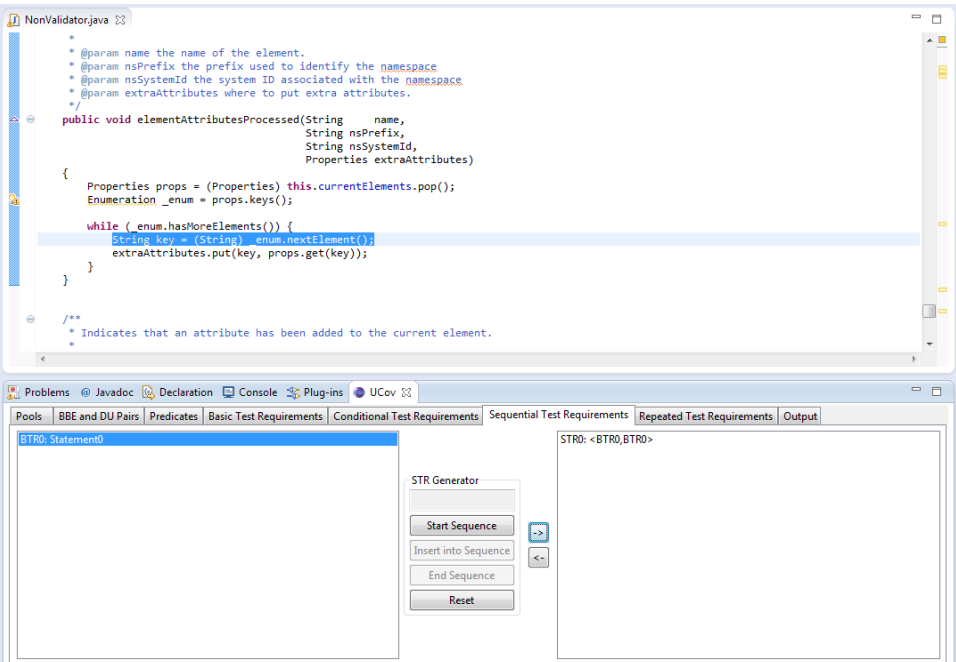

**Figure 13** – *str* specification in the *UCov* plugin

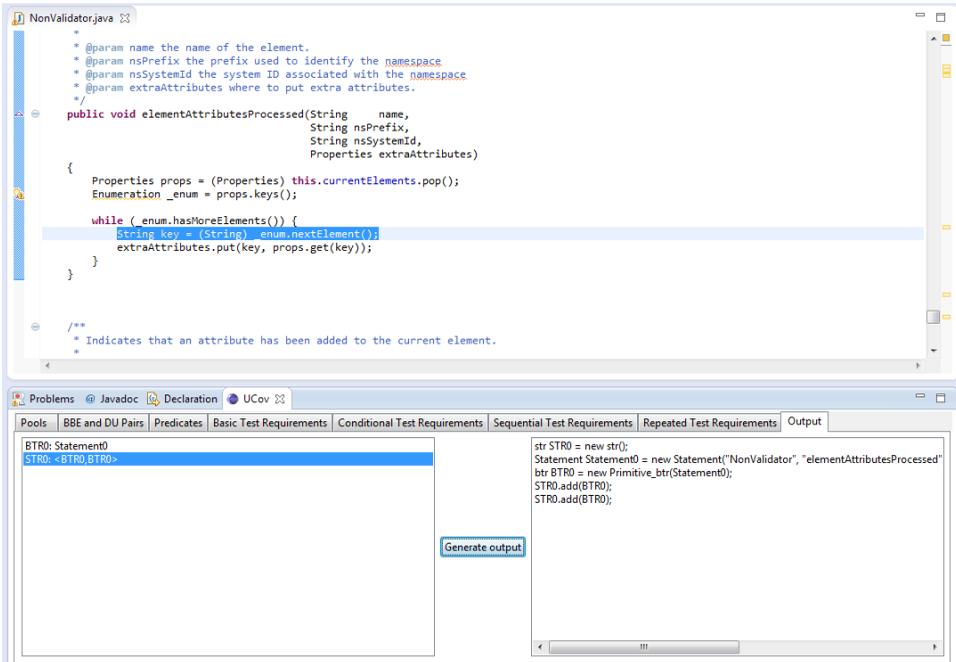

**Figure 14** – Generated code representing STR0

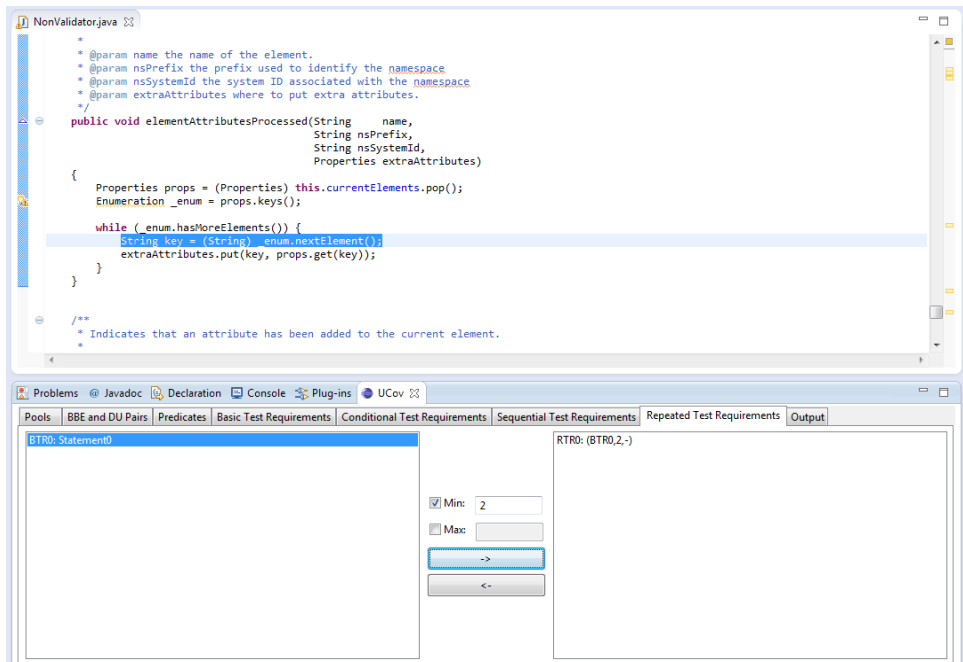

**Figure 15** – *rtr* specification in the *UCov* plugin

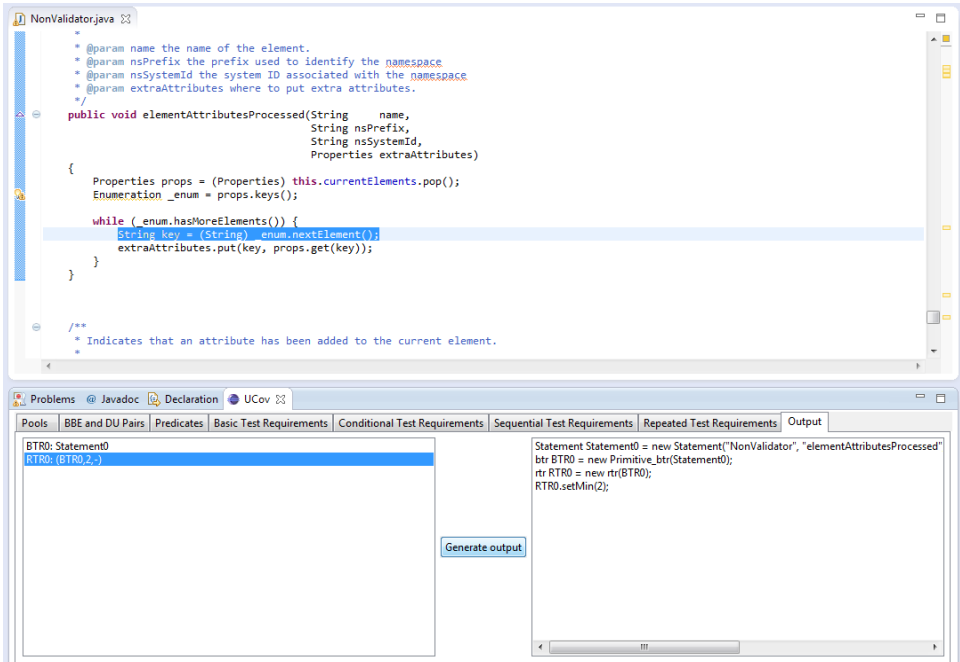

**Figure 16** – Generated code representing RTR0

*UCov* revealed that when executed in *NanoXML_v3*, the test case {Parser1_vw_v1.java, testvw_29.xml} did not actually exercise the bug fix (i.e., our user-defined test requirement was not covered), but instead resulted in an exception being thrown. Thus, in this real life case study, *UCov* alerted us that the test case associated with the bug fix became obsolete and that an alternate test case needs to be created.

To further investigate this case study, we manually tracked down the code change which rendered that test case obsolete and found out that it is related to the use of a different constructor of the *URL* class in method *openStream* in *StdXMLReader.java*. Noting that if the new constructor is replaced by the original one, $[[s_4]_{btr}, 2, \_]_{rtr}$ and $[<[s_4]_{btr}, [s_4]_{btr}>]_{str}$ would then be covered. The original code and the modified one are shown in Figure 17.

```
public Reader openStream(String publicID,
                         String systemID)
    throws MalformedURLException,
           FileNotFoundException,
           IOException
{
    systemID = "file:" + systemID;
    URL url = new URL(systemID);
    ...
```
(a)

```
public Reader openStream(String publicID,
                         String systemID)
    throws MalformedURLException,
           FileNotFoundException,
           IOException
{
    URL url = new URL(this.currentSystemID, systemID);
    ...
```
(b)

**Figure 17** –Code change that renders test case {Parser1_vw_v1.java, testvw_29.xml} obsolete. (a) *NanoXML_v1*; and (b) *NanoXML_v3*

## 5.2 TESTING SCENARIOS OF AN ALGORITHM

This case study targets the situation where a specific behavior needs to be tested. The application being considered is *tot_info*, one of the seven Siemens programs [Hutchins et al. 1994] that are widely used in the literature. More specifically, we inspect function *InfoTbl* that computes Kullback's information measure of a contingency table according to the following formula [Kullback 1968]:

$$\sum_{i=1}^{r}\sum_{j=1}^{c} x_{ij} \log(x_{ij}) - \sum_{i=1}^{r} x_i \log(x_i) - \sum_{j=1}^{c} x_j \log(x_j) + N \log N$$

where *r* and *c* are respectively the number of rows and columns in the contingency table, $x_{ij}$ is the value of the entry at row *i* and column *j*, $x_i$ is the sum of row *i*, $x_j$ is the sum of column *j*, and *N* is the sum of all entries in the table.

*InfoTbl* determines the information measure of a contingency table *T* by computing the four components of the formula above according to the pseudocode shown in Table 2. The algorithm starts by checking if *T* has at least two rows and two columns; if not, it returns -3 indicating that the table is too small. Lines 5-15 loop over the rows of *T*, compute the sum of each row and store it in array `xi`. At the same time, the sum `N` of all entries in the table is computed. If a negative entry is encountered during this process, the algorithm returns the "error" value -2. It also returns -1 if the total sum isn't strictly positive (lines 16-18). Similarly, the column sums are computed and stored in array `xj` (lines 19-25). The rest of the code computes each of the four components of the Kullback formula and aggregates the result in variable `info` as indicated in the table.

We distinguish three conditional checks in the code that prevent the algorithm from computing `log(0)`. Those are the ones at lines 28, 32, and 38. The first checks if the sum of the i$^{th}$ row is different than zero, the second checks if T[i,j] is different than zero, and the third checks if the sum of the j$^{th}$ column is different than zero. We argue that an important scenario to be covered is one in which the contingency table satisfies the following four conditions:

1) Is valid, i.e., has at least 2 rows and 2 columns, doesn't have negative entries, and isn't all zeros.
2) Has at least one row whose sum is zero.
3) Has at least one column whose sum is zero.
4) Has a strictly positive information measure so that a simple contingency table such as $\begin{bmatrix} 0 & 0 \\ 0 & 1 \end{bmatrix}$ with zero information measure would not be considered as a candidate.

We deem this scenario important because each of the three conditional checks on lines 28, 32, and 38 would evaluate both to *true* and *false* within the same test case. Applying *UCov*, the following test requirement, denoted by $\mathcal{P}$, captures the scenario at hand:

$$\mathcal{P} \equiv [<[[s_{13}]_{btr}, sum==0]_{ctr}, [[s_{24}]_{btr}, sum==0]_{ctr}, [[s_{42}]_{btr}, info>0]_{ctr}>]_{str}$$

For example, test cases based on the contingency table $\begin{bmatrix} 0 & 0 & 0 \\ 0 & 2 & 3 \\ 0 & 1 & 1 \end{bmatrix}$ satisfy $\mathcal{P}$ and thus are deemed important.

To verify whether *tot_info*'s original (full) test suite, which was downloaded from *SIR* (*sir.unl.edu*), covers $\mathcal{P}$, we created a modified version of *tot_info* in which we hard-coded some monitoring instructions that trigger a notification in case $\mathcal{P}$ is exercised. After running the modified version under the full test suite, we found out that no test case covers $\mathcal{P}$.

This real life case study shows that some scenarios that might be deemed important can go untested if not represented by non-generic test requirements such as those supported by *UCov*.

**Table 2** – Information Measure Algorithm

```
Input: Contingency table T, # rows r, # columns c
Output: -3 if r≤1 or c≤1, -2 if T contains a negative entry, -1 if T is
all zeros, Kullback measure of T otherwise

1.  if r≤1 OR c≤1
2.      return -3
3.  end if
4.  N = 0
5.  for i=1 to r
6.      sum = 0
7.      for j=1 to c
8.          if T[i,j] < 0
9.              return -2                    Computing row sums
10.         end if                                           and
11.         sum += T[i,j]                            total sum
12.     end for
13.     xi[i] = sum
14.     N += sum
15. end for
16. if N≤0
17.     return -1
18. end if
19. for j=1 to c
20.     sum = 0
21.     for i=1 to r
22.         sum += T[i,j]                 Computing column sums
23.     end for
24.     xj[j] = sum
25. end for
26. info = N×log(N)    /*** 4th component of Kullback's formula ***/
27. for i=1 to r
28.     if xi[i]>0
29.         info -= xi[i]×log(xi[i])      /*** 2nd component ***/
30.     end if
31.     for j=1 to c
32.         if T[i,j]>0
33.             info += T[i,j]×log(T[i,j])/*** 1st component ***/
34.         end if
35.     end for
36. end for
37. for j=1 to c
38.     if xj[j]>0
39.         info -= xj[j]×log(xj[j])      /*** 3rd component ***/
40.     end if
41. end for
42. return info
```

## 6. THREATS TO VALIDITY

*UCov* enables users to specify the intent of test cases. As with any specification task, the process is inevitably informal and subject to inaccuracies. However, with test case intent specification, the user does not start entirely from scratch as functional specification writers do, and is only required to associate an existing test case with existing program elements of concern. In addition, *UCov* facilitates this task by providing a programming interface supported by a user-friendly GUI plugin.

A major threat to the external validity of *UCov* is the fact that we were only able to present two real life case studies. Unfortunately, as it is the case for many newly proposed methodologies, we might not be able to measure the real effectiveness and usability of *UCov* until it gets extensively deployed and used by developers and testers.

We also recognize the following threats to the internal validity of *UCov*:
1) When specifying user-defined test requirements, the expressiveness of the Eclipse plugin or even the programming interface might not be adequate for the scenario at hand. Note that, currently, our plugin is as expressive as our programming interface.
2) The user might specify meaningless test requirements. We tried to remedy that within the Eclipse plugin by incrementally validating most steps taken by the user when specifying test requirements.
3) Cross referencing test requirements across versions might not lead to a perfect match in case the differences in code were considerable. We take a conservative approach to address this issue by asking the user to intervene in such cases.
4) The merit of *UCov* is that it complements existing structural coverage criteria. But in cases where the execution pattern characterizing the bug is simple such as a single statement, branch, or def-use, our approach might not have any value-added benefits since full coverage in existing criteria will suffice to reveal the bug. To illustrate this scenario, Appendix A walks through an example in which user-defined coverage as well as full statement coverage, both ensure that a bug fix is exercised.

## 7. RELATED WORK

Our first attempt for devising a methodology for test case intent preservation is described in Shaccour et al. [2013]. *UCov* addresses the shortcomings in that preliminary work, which are summarized below:
a) No user-friendly plugin or programming interface was provided to specify test requirements.
b) A user-defined test requirement was limited to a single def-use pair, which limited the applicability of the implementation.
c) Cross referencing test requirements across versions was very primitive, thus, requiring the intervention of the user in most cases.
d) No support for automated instrumentation to enable the checking for coverage of test requirements.

Testers leverage coverage criteria to maintain test suites that "hopefully" will: (1) exercise the functionality of the system under test (validation testing), (2) guard against previously detected/fixed defects (regression testing), and (3) increase the likelihood of detecting undiscovered defects (defect testing).

Over the years, researchers have proposed numerous coverage criteria many of which are discussed or listed in Ammann and Offutt [2008]. The fundamentals of data flow testing and def-use coverage were presented in [Laski and Korel 1983; Rapps and Weyuker 1985; Frankl and Weyuker 1988], and Harrold and Soffa [1994]. Data flow testing was contrasted against control flow and branch testing in Frankl and Weiss [1993] and Hutchins et al. [1994]. Coverage of logical expressions is treated in Ammann et al. [2003] and Jones and Harrold [2003]. Test case selection and prioritization is discussed in [Graves et al. 2001; Elbaum et al. 2002; Masri et al. 2007], and surveyed in Yoo and Harman [2012]. However, none of the above proposed techniques is capable of verifying or preserving the intent of test cases.

The *Rational PureCoverage* tool from IBM allows the tester through a GUI to restrict or focus the testing on select modules. Also here, test case intent cannot be verified or preserved.

Several techniques surveyed and compared in Yoo et al. [2013] aim at linking faults to test cases, and at ranking test cases based on their relevance to detected faults based on coverage metrics. These techniques employ statistical metrics and aim at fault localization. *UCov* differs in that it aims at establishing and *maintaining* the link between the fault, the test case, and the bug fix.

User defined coverage for hardware designs was introduced in Grinwald et al. [1998] as a methodology to annotate hardware logic written in VHDL or Verilog with coverage events. The method is not intended to preserve the intent of specific test cases and is limited to hardware designs. SystemVerilog [Bergeron 2005] supports a functional coverage specification language that introduces concepts like cover points, cover expressions, cover groups, and cross cover. Those coverage specifications are limited to hardware designs, are not related to specific test cases, and require knowledge of the whole design.

DSD-Crasher [Csallner 2008] aims at finding bugs by dynamically extracting invariants that describe the intended behavior of the program, excluding unwanted values from the domain of the program, exploring execution paths of the program that cover the invariants, and then generating test cases that cover the extracted paths. The work does not maintain the link between the detected invariants, the extracted paths, and the test cases. *UCov* can make use of the techniques proposed in DSD-Crasher to automatically extract execution paths and link them to existing test cases after the approval of the user.

## 8. CONCLUSIONS AND FUTURE WORK

Testers have relied on coverage criteria to assess the quality of test suites and to provide a stopping rule for testing. However, current criteria are based on coverage of simple and generic program elements such as statements, branches, and def-use pairs which in most case cannot characterize non-trivial behaviors or specific behaviors deemed critical for testing. To address this issue, we present *UCov*, a methodology and tool for precise test case intent verification in regression test suites. *UCov* complements existing coverage criteria by focusing the testing on important code patterns or behaviors that could go untested otherwise. That is, *UCov* allows the tester to specify user-defined test requirements to be covered; it also facilitates test case intent preservation.

As part of future work, we intend to:
1) Enhance our plugin by adding new functionality that improves its usability.
2) Conduct experiments involving real users to assess the effectiveness of *UCov*.
3) Fully support test case intent preservation. That is, in case of a failed test intent verification, automated test case generation will be performed whose aim is to satisfy the user-defined test requirement and thus preserve the intent of the test case.
4) Investigate extracting test requirements associated with bug fixes automatically from source control repositories.

# APPENDIX A

## Example with a Simple User-Defined Test Requirement

Method *isPrime(int x)* is meant to return *true* if *x* is a prime number, and *false* otherwise. $P_1$ is a faulty implementation of *isPrime(int x)* where the bug is in statement $s_0$.

```
// P1
public static boolean isPrime(int x) {
      if (x <= 1) return false;
      else if (x == 2) return true;
      else {
        int UpperLimit = (int) (Math.sqrt (x) +1);
        for(int divisor = 2 ; divisor <= UpperLimit ; divisor++ ){
s0          if(x % divisor != 0) { // bug: should be ==
s1              return false;
            }
        }
   s2   return true;
      }
}
```

Assume that we are set to incrementally build a test suite $T$ that achieves full statement coverage. Starting with x = 1, we select 6 test cases that cumulatively cover all the statements in $P_1$, namely, $t_1$:{1, false}, $t_2$:{2, true}, $t_3$:{3, true}, $t_4$:{4, false}, $t_5$:{5, true}, and $t_6$:{6, false}. Note how due to the bug at $s_0$, $t_3$ and $t_6$ return unexpected values, i.e., they are failing test cases. And since $t_4$ and $t_5$ do not increase coverage, they are ignored, thus, leading to $T_1$ = {$t_1$, $t_2$, $t_3$, $t_6$}, which yields full statement coverage.

As a result of $t_3$ and $t_6$ the bug is revealed and fixed in $P_2$. Also, applying *UCov*, $t_3$ is coupled with test requirement $[<[s_0]_{btr}, [s_2]_{btr}>]_{str}$ and $t_6$ is coupled with test requirement $[<[s_0]_{btr}, [s_1]_{btr}>]_{str}$.

```
// P2
public static boolean isPrime(int x) {
      if (x <= 1) return false;
      else if (x == 2) return true;
      else {
        int UpperLimit = (int) (Math.sqrt (x) +1);
        for(int divisor = 2 ; divisor <= UpperLimit ; divisor++ ){
s0          if(x % divisor == 0) { // bug is fixed
s1              return false;
            }
        }
   s2   return true;
      }
}
```

In case $P_2$ is refactored into $P_3$ shown below, *UCov* detects that the intents of $t_3$ and $t_6$ are not preserved anymore, since $s_0$ is not executed in either case.

Consequently, the user would replace $t_3$ by $t_7 = \{7, \text{true}\}$ which covers $\{s_0, s_2\}$, and $t_6$ by $t_9 = \{9, \text{false}\}$ which covers $\{s_0, s_1\}$. Now the test suite becomes $T_2 = \{t_1, t_2, t_7, t_9\}$ as opposed to $T_1 = \{t_1, t_2, t_3, t_6\}$. Note how if the user kept $T_1$, $s_0$ and $s_1$ (and the bug fix) would not be exercised.

Alternatively, if instead of applying *UCov*, the tester tried to achieve full statement coverage, she would realize that $T_1 = \{t_1, t_2, t_3, t_6\}$ is deficient for $P_3$, and that any test suite that achieves full coverage, would actually cover the bug fix. That is, in this example, full statement coverage is as effective as *UCov*.

```
// P3
public static boolean isPrime(int x) {
        if (x <= 1) return false;
        else if (x == 2) return true;
        else if (x % 2 == 0)    return false; // Added Code
        else {
          int UpperLimit = (int) (Math.sqrt (x) +1);
          for(int divisor=3 ; divisor <= UpperLimit ; divisor+=2){ // modified code
    s0         if(x % divisor == 0) {
    s1              return false;
                }
          }
    s2   return true;
       }
}
```